\documentclass[aps,prl,twocolumn,showpacs,byrevtex]{revtex4}
\let\oldhat\hat
\renewcommand{\hat}[1]{\oldhat{\mathbf{#1}}}

\usepackage{times,mathptm}% for making `pdf' file
\usepackage[dvips]{graphicx,color}% for inclusion of graphics file(s)

\begin{document}
\title{Coupling between Charge, Lattice, Orbital, and Spin in a Charge Density Wave of 1$T$-TaS$_2$}
\author{Seho Yi$^1$, Zhenyu Zhang$^2$, and Jun-Hyung Cho$^{1,2*}$}
\affiliation{$^1$ Department of Physics and Research Institute for Natural Sciences, Hanyang University, Seoul 133-791, Korea\\
$^2$ ICQD, Hefei National Laboratory for Physical Sciences at the Microscale,
and Synergetic Innovation Center of Quantum Information and Quantum Physics,
University of Science and Technology of China, Hefei, Anhui 230026, China
}
\date{\today}

\begin{abstract}
Two-dimensional layered transition metal dichalcogenide (TMDC) materials often exhibit exotic quantum matter phases due to the delicate coupling and competitions between the charge, lattice, orbital, and spin degrees of freedom. Surprisingly, we here present, based on first-principles density-functional theory calculations, the incorporation of all such degrees of freedom in a charge density wave (CDW) of monolayer (ML) TMDC 1$T$-TaS$_2$. We reveal that the CDW formed via the electron-phonon coupling is significantly stabilized by the orbital hybridization. The resulting lattice distortion to the ``David-star" superstructure constituted of one cental, six nearest-neighbor, and six next-nearest-neighbor Ta atoms is accompanied by the formation of quasimolecular orbitals due to a strong hybridization of the Ta $t_{\rm 2g}$ orbitals. Furthermore, the flat band of the quasimolecular orbital at the Fermi level has a spin splitting caused by an intramolecular exchange, yielding a full spin polarization with a band-gap opening. Our finding of the intricate charge-lattice-orbital-spin coupling in ML 1$T$-TaS$_2$ provides a framework for the exploration of various CDW phases observed in few-layer or bulk 1$T$-TaS$_2$.

\end{abstract}
\pacs{71.20.Ps, 73.22.-f, 81.07.-b}
\maketitle
%\begin{multicols}{2}

%\vspace{0.4cm}
%\section{I. INTRODUCTION}
%\vspace{0.4cm}

Low-dimensional electron systems realized in layered 2D materials have attracted much attention in contemporary condensed-matter physics because they tend to exhibit an interesting variety of quantum matter phases, such as charge density wave (CDW)~\cite{Shen,Xi,Keum,chen}, magnetism~\cite{Ma}, and superconductivity~\cite{Fri}. These macroscopic quantum condensates are associated with the interactions of quasi-particles that involve the charge, lattice, orbital, and spin degrees of freedom. Specifically, the CDW formation driven by the Fermi surface nesting~\cite{Peierls} or the strong coupling between an electron charge modulation and a periodic lattice distortion~\cite{Johannes,Zhua} has been widely observed in metallic layered transition metal dichalcogenide (TMDC) materials with van der Waals interlayer interactions~\cite{Shen,Xi,Keum,chen}.

As a prototype of layered TMDCs, bulk 1$T$-TaS$_2$ exhibits a variety of CDW phases including an incommensurate CDW phase below 550 K, a nearly commensurate CDW phase below 350 K, and a commensurate CDW phase below 190 K. Especially, a nearly commensurate CDW phase is transformed into a commensurate CDW one with accompanying a metal-insulator transition~\cite{Sip,Salvo,Fazekas}. The latter CDW phase has been known to be a 2D Mott insulator~\cite{Fazekas,Wilson,Kim,Perfetti,Perfetti2,Yeom}. According to this CDW-driven Mott scenario, the CDW is also accompanied by the so-called David-star (DS) distortion within a large unit cell of ${\sqrt{13}}{\times}{\sqrt{13}}$ [see Fig. 1(a)], which substantially reduces the bandwidth at the Fermi level $E_{\rm F}$ to invoke a Mott insulator with the enhanced on-site electron-electron interaction $U$. This scenario features the generation of two types of gaps: i.e., the CDW gap ${\Delta}_{\rm CDW}$ and Mott gap ${\Delta}_{\rm Mott}$~\cite{Yeom}. By contrast, a recent combined angle-resolved photoemission spectroscopy (ARPES) and density-functional theory (DFT) study reported that the CDW within the 2D TaS$_2$ layers involves the complex orbital textures, which play a crucial role in determining the band dispersion and gap structure through their interlayer interactions~\cite{Ritschel}. Meanwhile, the DFT calculation without including an on-site $U$ showed that the CDW in monolayer (ML) 1$T$-TaS$_2$ produces a gap opening caused solely by spin polarization~\cite{zhang2014prb}, implying that this gap seems not to be related with Mott physics alone. Therefore, despite many theoretical and experimental studies~\cite{Sip,Salvo,Fazekas,Wilson,Kim,Perfetti,Perfetti2,Yeom,Ritschel,zhang2014prb}, the microscopic nature of the CDW as well as the origin of the gap openings in 1$T$-TaS$_2$ are still veiled. It is, however, more likely that the CDW formation would involve the intricate interplays among the charge, lattice, orbital, and spin degrees of freedom.

\begin{figure}[h!t]
\includegraphics[width=8cm]{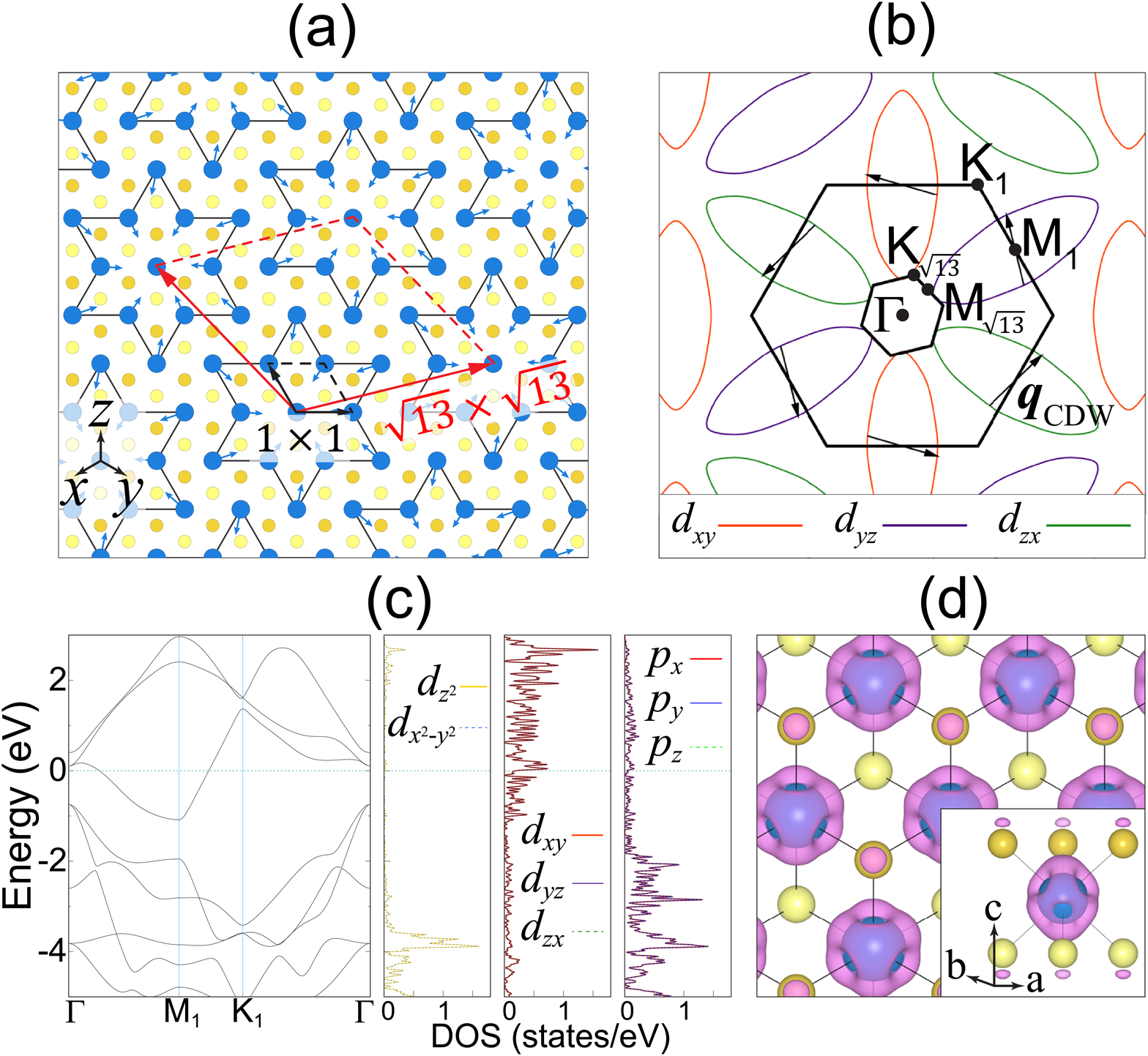}
\caption{(Color online) (a) Optimized 1${\times}$1 structure of ML 1$T$-TaS$_2$. The DS distortion with the ${\sqrt{13}}{\times}{\sqrt{13}}$ periodicity is indicated by the arrows, with the pattern of the DS. The large and small (dark and bright) circles represent Ta and S (upper and lower) atoms, respectively. For the analysis of orbital characters, we choose the $x$, $y$, and $z$ axes pointing nearly along the three Ta$-$S bonds. The Brillouin zones of the 1${\times}$1 and ${\sqrt{13}}{\times}{\sqrt{13}}$ structures are drawn in (b). The ellipses represent the Fermi surface with $q_{\rm CDW}$. The 1${\times}$1 band structure together with the partial DOS projected onto the Ta 5$d$ and S 3$p$ orbitals is shown in (c). The charge character of the $t_{\rm 2g}$ states, which is integrated over an energy window between $E_{\rm F}-$0.15 and $E_{\rm F}$+0.15 eV, is displayed in (d). In the inset, the side view is drawn with the lattice unit vectors $a$, $b$, and $c$.}
\label{figure:1}
\end{figure}

Here, we focus on the microscopic nature of a CDW in an isolated ML, which is a basic building block of the layered structure of 1$T$-TaS$_2$. Recently, the micromechanical exfoliation technique can realize ultra-thin films of 1$T$-TaS$_2$ with ML~\cite{Albert} and few-layer thicknesses~\cite{Tsen}. Based on first-principles DFT calculations, we find that the DS distortion takes place barrierlessly with a strong orbital hybridization. As the lattice distortion increases, the Ta $t_{\rm 2g}$ orbitals hybridize with each other to form the bonding (B), nonbonding (NB), and antibonding (AB) quasimolecular orbitals (QMOs) around $E_{\rm F}$. This lattice-orbital coupling significantly stabilizes the CDW with the gap openings (corresponding to ${\Delta}_{\rm CDW}$) between the QMOs. Specifically, the NB QMO features a quasilocalized electron around the central Ta atom in the DS unit, forming a flat band at $E_{\rm F}$. We reveal that this half-filled band is spin-polarized due to the local polarization effect caused by an intramolecular exchange (i.e., Stoner parameter $I$), giving rise to a gap opening corresponding to ${\Delta}_{\rm Mott}$~\cite{severin-prl93}. The results elucidate that the underlying mechanism of the gap openings in the CDW phase of ML 1$T$-TaS$_2$ is associated with the intricate charge-lattice-orbital-spin coupling. This complex, unique picture offers the framework in which the still perplexing features of various CDW phases and their transitions in few-layer or bulk 1$T$-TaS$_2$ have to be understood.

Our DFT calculations were performed using the the Vienna $ab$ initio simulation package (VASP) code with the projector-augmented wave (PAW) method~\cite{vasp1,vasp2}. For the treatment of exchange-correlation energy, we employed the generalized-gradient approximation functional of Perdew-Burke-Ernzerhof (PBE)~\cite{pbe} and the hybrid functional of Heyd-Scuseria-Ernzerhof (HSE)~\cite{hse}. The ML 1$T$-TaS$_2$ system was modeled by a periodic slab geometry with ${\sim}$25 {\AA} of vacuum in between the slabs. A plane wave basis was employed with a kinetic energy cutoff of 400 eV, and the ${\bf k}$-space integration was done with the 21${\times}$21 and 6${\times}$6 meshes in the Brillouin zones of 1${\times}$1 and ${\sqrt{13}}{\times}{\sqrt{13}}$, respectively. All atoms were allowed to relax along the calculated forces until all the residual force components were less than 0.005 eV/{\AA}.

We begin to optimize the 1${\times}$1 structure of ML 1$T$-TaS$_2$ using the PBE calculation. The optimized 1${\times}$1 structure and its Brillouin zone are displayed in Fig. 1(a) and 1(b), respectively. Figure 1(c) shows the calculated band structure and partial density of states (DOS) of the 1${\times}$1 structure. There are three bands around $E_{\rm F}$, which originate mostly from the Ta $d_{xy}$, $d_{yz}$, and $d_{zx}$ orbitals. Meanwhile, the Ta $d_{z^2}$ and $d_{x^2-y^2}$ states are located at around 3.8 eV below $E_{\rm F}$, which overlap with the S $p_{x}$, $p_{y}$, and $p_{z}$ orbitals. Therefore, Ta $d$ states are split into the upper $t_{\rm 2g}$ states with a large band width of 4.05 eV [see Fig. 1(c)] and the lower $e_{\rm g}$ states that participate in covalent bonds with S atoms. We find that the lowest $t_{\rm 2g}$ state crosses $E_{F}$, indicating a metallic feature. As shown in Fig. 1(d), the charge character of the $t_{2g}$ states near $E_{\rm F}$ represents a strong electron accumulation around Ta atoms with a negligible amount of charges at S atoms. This weak hybridization of $t_{\rm 2g}$ with the S 3$p$ orbitals indicates a direct nearest neighbors hopping between $t_{2g}$ orbitals, which can be enhanced by the DS distortion. This feature leads to the formation of QMOs caused by the lattice-orbital coupling, as discussed below.

As shown in Fig. 1(b), the Fermi surface of the 1${\times}$1 structure consists of six elliptically shaped electron pockets around the $M_{\rm 1}$ points, in good agreement with previous DFT calculations~\cite{Myron,Shao} and ARPES measurements~\cite{Pillo,clerc2006prb}. It is noted that the Peierls-type CDW formation is driven by an electron charge modulation due to Fermi surface nesting~\cite{Peierls,Johannes,Zhua}. The present nesting vector $q_{\rm CDW}$ corresponding to the periodic ${\sqrt{13}}{\times}{\sqrt{13}}$ lattice distortion is drawn in Fig. 1(b), connecting flat parts of the Fermi surface~\cite{Myron,Pillo}. However, this topology of the Fermi surface was reported to produce weak peaks around $q_{\rm CDW}$ in the imaginary part of the static electronic susceptibility, implying that Fermi surface nesting is not a plausible explanation for the CDW instability~\cite{zhang2014prb}. Instead, it has been accounted for in terms of the electron-phonon coupling (EPC)~\cite{zhang2014prb,clerc2006prb}. However, the driving force of how this EPC occurs is still missing. We will show below that the EPC-driven CDW is significantly stabilized by the formation of QMOs with a strong lattice-orbital coupling. Interestingly, as shown in Fig. 1(b), each electron pocket at the Fermi surface exhibits an identical orbital character among the degenerate $d_{xy}$, $d_{yz}$, and $d_{zx}$ orbitals. Note that the electron pockets in the opposite ${\bf k}$ directions have the same orbital character. As the lattice distortion increases, this ${\bf k}$-dependent orbital character easily induces a hybridization of the $d_{xy}$, $d_{yz}$, and $d_{zx}$ orbitals (see Fig. S1 of the Supplemental Material~\cite{SM}). Thus, we can say that the EPC is accompanied with the lattice-orbital coupling to form the CDW in ML 1$T$-TaS$_2$.

\begin{figure}[h!t]
\includegraphics[width=8cm]{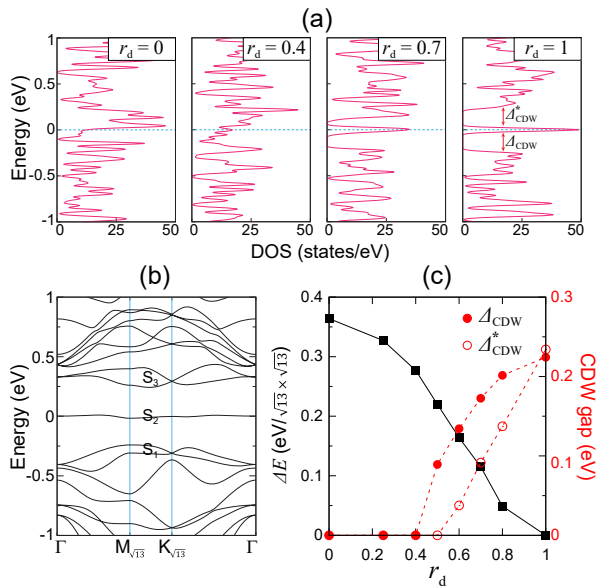}
\caption{(Color online) (a) Calculated total DOS with respect to $r_d$ and (b) band structure obtained at $r_d$ = 1. The energy zero represents the Fermi level. In (c), the total-energy variation relative to the DS structure ($r_d$ = 1) is given as a function of $r_d$. The variations of ${\Delta}_{\rm CDW}$ and ${\Delta}^{*}_{\rm CDW}$ are also given in (c).}
\label{figure:2}
\end{figure}

\begin{figure*}[h!t]
\includegraphics[width=16cm]{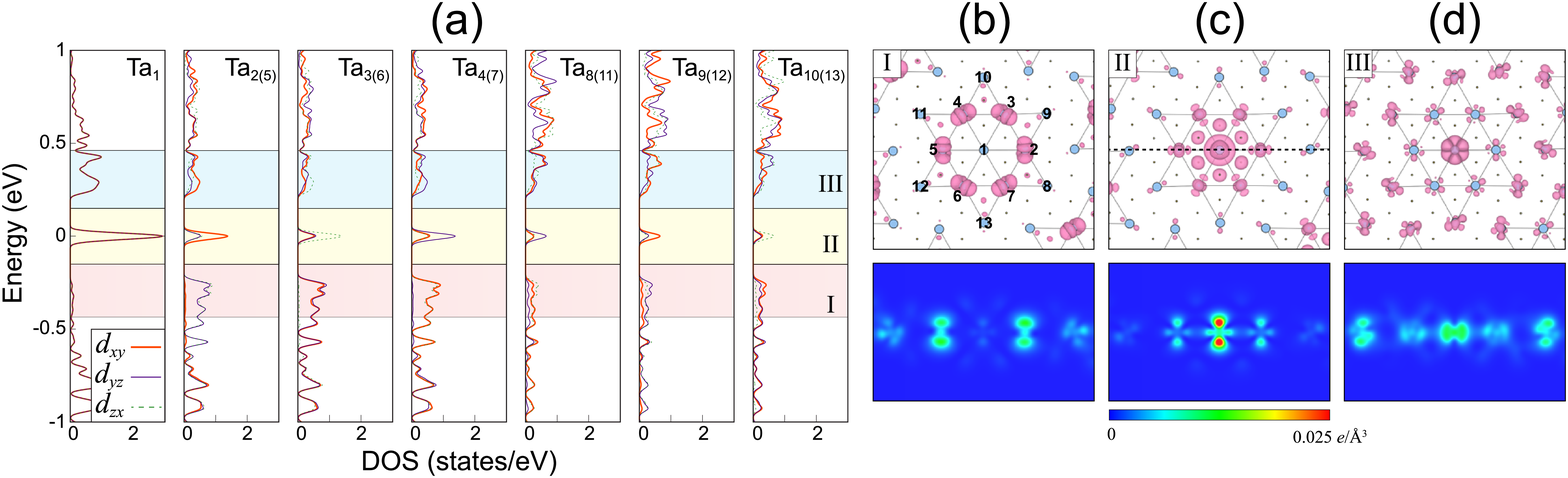}
\caption{(Color online) (a) Calculated partial DOS projected onto thirteen Ta atoms in the DS unit. The labeling of each Ta atom is given in (b). The top views of the charge densities ${\rho}_{\rm I}$, ${\rho}_{\rm II}$, and ${\rho}_{\rm III}$, integrated over the three energy windows I, II, and III, are displayed in (b), (c), and (d), respectively. Here, ${\rho}_{\rm I}$ and ${\rho}_{\rm III}$ are drawn with the isosurface of 0.006 $e$/{\AA}$^3$, while ${\rho}_{\rm II}$ with 0.003 $e$/{\AA}$^3$. The side view (in lower panel) is drawn in the vertical plane along the dashed line in (c).}
\label{figure:3}
\end{figure*}

To explore how the electronic states change during the CDW formation, we calculate the band structure and DOS as a function of the lattice distortion, defined as the fractional ratio $r_d$ = ${\Delta}d$/$d_0$. Here, ${\Delta}d$ represents the displacement of a nearest-neighbor (or next-nearest neighbor) Ta atom from the 1${\times}$1 structure, with $d_0$ referring to the full displacement obtained at the DS structure. The present value of $d_0$ for the nearest-neighbor (next-nearest neighbor) Ta atoms is 0.21 (0.25) {\AA}, in good agreement with previous DFT calculations~\cite{zhang2014prb}. The calculated $d$-orbitals projected band structure is given as a function of $r_d$ in Fig. S2 of the Supplemental Material~\cite{SM}, while its total DOS in Fig. 2(a). We find that the hybridization of the $d_{xy}$, $d_{yz}$, and $d_{zx}$ orbitals increases with increasing the lattice distortion. As a consequence of this lattice-orbital coupling, it is seen in Figs. 2(a) and 2(b) that, when $r_d$ $>$ ${\sim}$0.4, the $t_{\rm 2g}$ states are split into the $S_1$ (doublet), $S_2$, and $S_3$ (doublet) states near $E_{\rm F}$. Specifically, at $r_d$ = 1 (DS structure), the band gaps ${\Delta}_{\rm CDW}$ and ${\Delta}^{*}_{\rm CDW}$ become 0.22 and 0.23 eV [see Figs. 2(b) and 2(c)], respectively. According to the scanning tunneling spectroscopy (STS) experiment of Cho $et$ $al$.~\cite{Yeom}, two prominent peaks were observed at $-$0.19 and +0.23 eV, which were interpreted as the lower and upper Hubbard bands, respectively. Beyond such a Mott gap ${\Delta}_{\rm Mott}$ ${\sim}$ 0.42 eV, an additional peak was observed at around $-$0.30 eV, giving rise to ${\Delta}_{\rm CDW}$ ${\sim}$ 0.30 eV~\cite{Yeom}. This CDW gap is well comparable with our prediction of ${\Delta}_{\rm CDW}$.

Figure 3(a) shows the partial DOS projected onto thirteen Ta atoms in the DS unit. The charge densities ${\rho}_{\rm I}$, ${\rho}_{\rm II}$, and ${\rho}_{\rm III}$, integrated over the three energy windows I (between $E_{\rm F}-$0.45 and $E_{\rm F}-$0.15 eV), II (between $E_{\rm F}-$0.15 and $E_{\rm F}$+0.15 eV), and III (between $E_{\rm F}$+0.15 and $E_{\rm F}$+0.43 eV), are displayed in Figs. 3(b), 3(c), and 3(d), respectively. Note that these energy windows I, II, and III separately cover the $S_1$, $S_2$, and $S_3$ states [see Fig. 2(b)], respectively. We find that the partial DOS and charge characters of the $S_1$, $S_2$, and $S_3$ states represent the B, NB, and AB QMOs, respectively: i.e., (i) ${\rho}_{\rm I}$ exhibits a charge delocalization encircling the six nearest-neighbor Ta atoms, (ii) ${\rho}_{\rm II}$ has a quasilocalization around the central Ta atom, and (iii) ${\rho}_{\rm III}$ is evenly distributed over the thirteen Ta atoms. Such characters of the QMOs can be seen as $r_d$ increases (see Fig. S3 of the Supplemental Material~\cite{SM}), indicating that the lattice distortion-induced hybridization of the $t_{\rm 2g}$ states produces the B, NB, and AB QMOs. Specifically, the quasilocalized charge character of the NB QMO [see Fig. 3(c)] agrees well with a previous DFT result~\cite{Ritschel} that the electron density of the highest occupied state of bulk 1$T$-TaS$_2$ exhibits a complex orbital texture pointing along the $c$ axis. Since the total energy decreases monotonically up to ${\sim}$0.36 eV per DS unit at $r_d$ = 1 [see Fig. 2(c)], we can say that the EPC-driven CDW is significantly stabilized by the formation of the B, NB, and AB QMOs.

\begin{figure}[h!t]
\includegraphics[width=8cm]{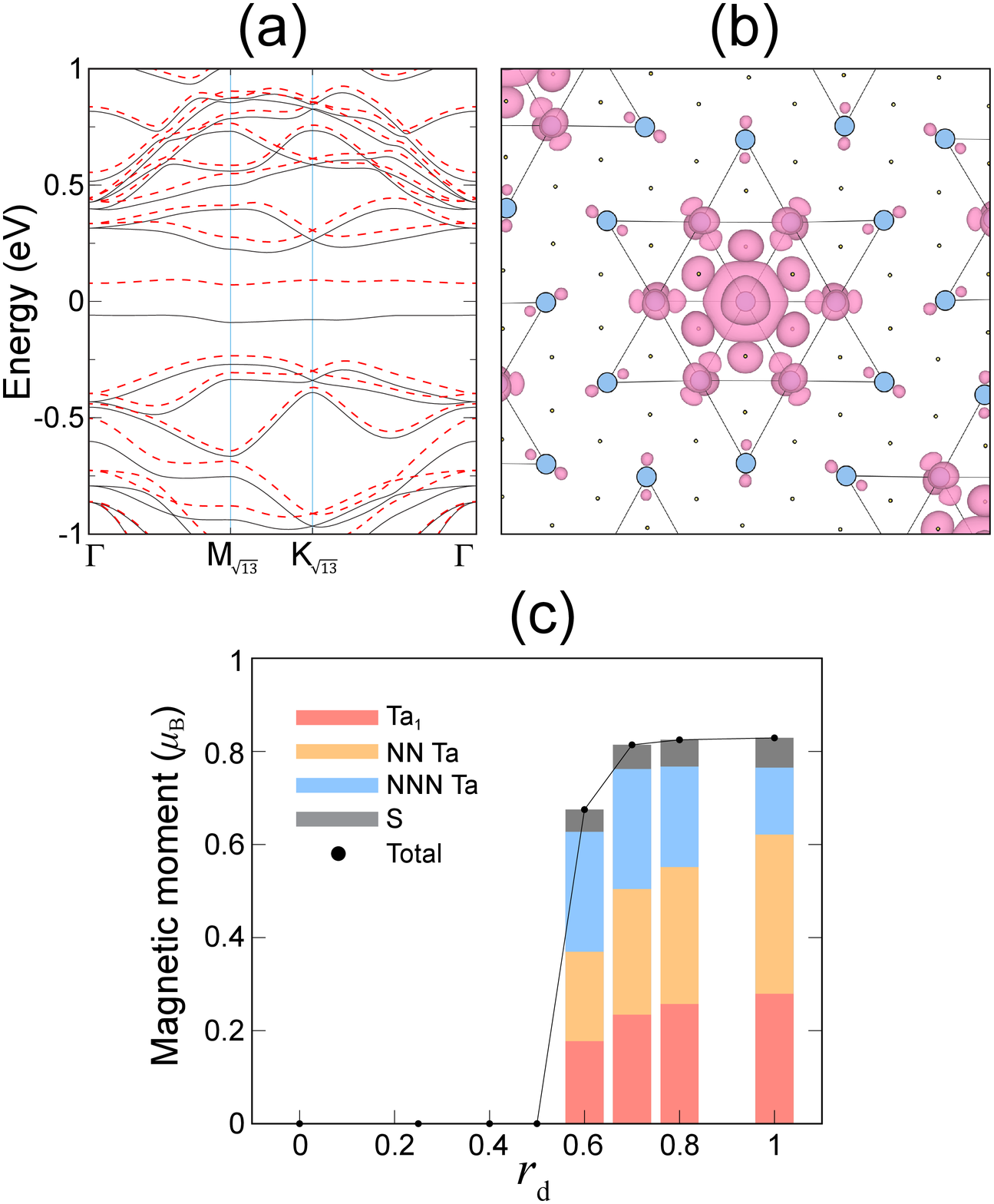}
\caption{(Color online) (a) Calculated band structure and (b) spin density distribution of the FM state obtained at $r_d$ = 1. The majority and minority bands in (a) are drawn with the solid and dashed lines, respectively. The spin density in (b) is displayed with an isosurface of 0.0015 $e$/{\AA}$^3$. The spin moments arising from the central, nearest-neighbor (NN), and next-nearest-neighbor (NNN) Ta atoms and S atoms are decomposed in (c). }
\label{figure:3}
\end{figure}

It is noticeable that the NB QMO has a half-filled band with a very narrow bandwidth of 0.03 eV [see Fig. 2(b)]. Due to such a flat-band character of the NB QMO, the spin-polarized configuration corresponding to the ferromagnetic (FM) state is found to be more stable than the non-spin-polarized one by 42.5 meV per DS unit. The band structure of this FM state is displayed in Fig. 4(a), showing a band-gap opening of 0.15 eV with a full spin polarization. As shown in Fig. 4(b), the spin density of the FM state is distributed around the cental Ta atom (labeled as Ta$_1$). We note that the charge character of the NB QMO represents a quasilocalized electron that is not only localized at Ta$_1$ but also some distributed over the DS cluster centered at Ta$_1$ [see Fig. 3(c)]. Such a charge distribution of the NB QMO in ML 1$T$-TaS$_2$ is found to be almost the same as that of the highest occupied molecular orbital (HOMO) of an isolated DS molecule (see Fig. S4 of the Supplemental Material~\cite{SM}). This implies that ML 1$T$-TaS$_2$ has a weak electronic coupling between neighboring DS clusters. Considering the fact that the ground state of an isolated DS molecule is spin polarized (see Fig. S4 of the Supplemental Material~\cite{SM}), we can say that the spin splitting of the NB QMO in ML 1$T$-TaS$_2$ is attributed to the local polarization effect caused by an intramolecular exchange (i.e., Stoner parameter $I$). It is noted that, if the HSE functional with a mixing factor of ${\alpha}$ = 0.125 controlling the amount of exact Fock exchange energy is employed~\cite{hse}, the spin-polarization gap in ML 1$T$-TaS$_2$ increases to 0.35 eV (see Fig. S5 of the Supplemental Material~\cite{SM}), close to ${\Delta}_{\rm Mott}$ = 0.42 eV measured~\cite{Yeom} from 1$T$-TaS$_2$. Therefore, we can say that the gap opening of ${\Delta}_{\rm Mott}$ is most likely to be associated with an intramolecular exchange interaction within the DS cluster. Indeed, in the spin-polarized DFT scheme, it was known that $I$ representing intramolecular-exchange or self-exchange energy is the on-site Coulomb interaction $U$~\cite{severin-prl93}.

Figure 4(c) shows that the spin polarization of ML 1$T$-TaS$_2$ appears from $r_d$ $>$ 0.5, where both ${\Delta}_{\rm CDW}$ and ${\Delta}^{*}_{\rm CDW}$ begin to open [see Fig. 2(c)]. This indicates that, as the flat band of the NB QMO is created, its full spin polarization takes place simultaneously. Therefore, all the FM states in $r_d$ $>$ 0.5 have a total spin moment of 1 ${\mu_{\rm B}}$ per DS unit. However, since the spin moment is calculated by integrating the spin density inside the PAW sphere with a radius of 1.62 (1.03) {\AA} for Ta (S), the sum of the spin moments of all atoms is less than 1 ${\mu_{\rm B}}$ [see Fig. 4(c)]. Interestingly, it is seen that, as $r_d$ increases, the spin moments of the central and nearest-neighbor Ta atoms increase while those of the next-nearest-neighbor Ta atoms decrease, reflecting more localization around the central Ta atom.

To examine how large intermolecular exchange interaction exists in ML 1$T$-TaS$_2$, we perform an additional PBE calculation for the antiferromagnetic (AFM) state with twice larger unit cell. We find that the AFM state is less stable than the FM one by 1.9 meV per DS unit. As shown in Fig. S6 of the Supplemental Material~\cite{SM}, the AFM state has a band gap of 0.12 eV, smaller than that (0.15 eV) of FM. This indicates that the AFM spin ordering gives a relatively lower electronic energy gain, compared to the FM one. From the energy difference between the AFM and FM states, the exchange coupling constant $J$ between the magnetic moments of neighboring DS clusters is estimated to be as small as ${\sim}$2 meV. This weak intermolecular exchange interaction is consistent with the experimental observation of low-temperature paramagnetism in 1$T$-TaS$_2$~\cite{para}.

Although the present results are based on a single ML of 1$T$-TaS$_2$, the proposed formation mechanisms of ${\Delta}_{\rm CDW}$ and ${\Delta}_{\rm Mott}$ will be useful for understanding the CDW phase of bulk 1$T$-TaS$_2$. It is noteworthy that the energy gain caused by ${\Delta}_{\rm CDW}$ formation [see Fig. 2(c)] is about nine times larger than that (${\sim}$42.5 meV) for ${\Delta}_{\rm Mott}$ formation, indicating that the CDW is much more stable against perturbation compared to the spin polarization. Especially, ${\Delta}_{\rm Mott}$ generated by the spin splitting of the NB QMO may be very susceptible to the interlayer interactions which can be varied by changing the layer stacking. Indeed, a recently combined ARPES and DFT study~\cite{Ritschel} showed that the bands (relevant to ${\Delta}_{\rm Mott}$) near $E_{\rm F}$ are strongly affected by different layer stackings due to their variation of the interlayer hybridization with the complex orbital textures, giving rise to metallic and semiconducting states. On the other hand, the bands (relevant to ${\Delta}_{\rm CDW}$) away from $E_{\rm F}$ remain intact. Therefore, taking into account the presently proposed QMOs and spin polarization in the CDW formation, more accurate understanding of the effect of the interlayer interactions on ${\Delta}_{\rm CDW}$ and ${\Delta}_{\rm Mott}$ will be an interesting subject for future work.

In summary, we have performed a first-principles DFT study to explore the microscopic nature of a CDW in ML 1$T$-TaS$_2$. We found that the DS distortion within the ${\sqrt{13}}{\times}{\sqrt{13}}$ unit cell is accompanied with a strong hybridization of the Ta $t_{\rm 2g}$ orbitals, giving rise to the formation of the B, NB, AB QMOs. This lattice-orbital coupling significantly stabilizes the CDW formation with a gap opening of ${\Delta}_{\rm CDW}$. Furthermore, the flat band of the NB QMO is spin-polarized due to the local polarization effect caused by an intramolecular exchange, opening a gap corresponding to ${\Delta}_{\rm Mott}$. Therefore, we demonstrated the presence of the intricate charge-lattice-orbital-spin coupling behind the CDW formation in ML 1$T$-TaS$_2$. This complex, unique feature of the 2D CDW phase is anticipated to shed new light on various CDW phases and their transitions in few-layer or bulk 1$T$-TaS$_2$, the microscopic nature of which has been elusive.

\vspace{0.4cm}

%\vspace{0.4cm}
%\noindent {\bf Acknowledgement.}
%\vspace{0.4cm}
This work was supported by National Research Foundation of Korea (NRF) grant funded by the Korean Government (Nos. 2015R1A2A2A01003248 and 2015M3D1A1070639). Z.Z is supported by National Natural Science Foundation of China (Nos. 11634011 and 61434002). The calculations were performed by KISTI supercomputing center through the strategic support program (KSC-2016-C3-0059) for the supercomputing application research.

                  %%%%%  REFERENCES  %%%%%
\noindent $^{*}$ Corresponding author: chojh@hanyang.ac.kr

%\centering
%\includegraphics{fig1}

\widetext
\clearpage

\vspace{2.4cm}
\centering{{\huge Supplemental materials for "Coupling between Charge, Lattice, Orbital, and Spin in a Charge Density Wave of 1$T$-TaS$_2$}}

\vspace{0.8cm}

\makeatletter
\renewcommand{\fnum@figure}{\figurename~S\thefigure}
\renewcommand{\fnum@table}{\tablename~S\thetable}
\setcounter{figure}{0}
\makeatother

\vspace{1.4cm}
{\bf \large 1. Orbital characters of the band crossing the Fermi level}

\begin{figure}[h]
\includegraphics[width=16cm]{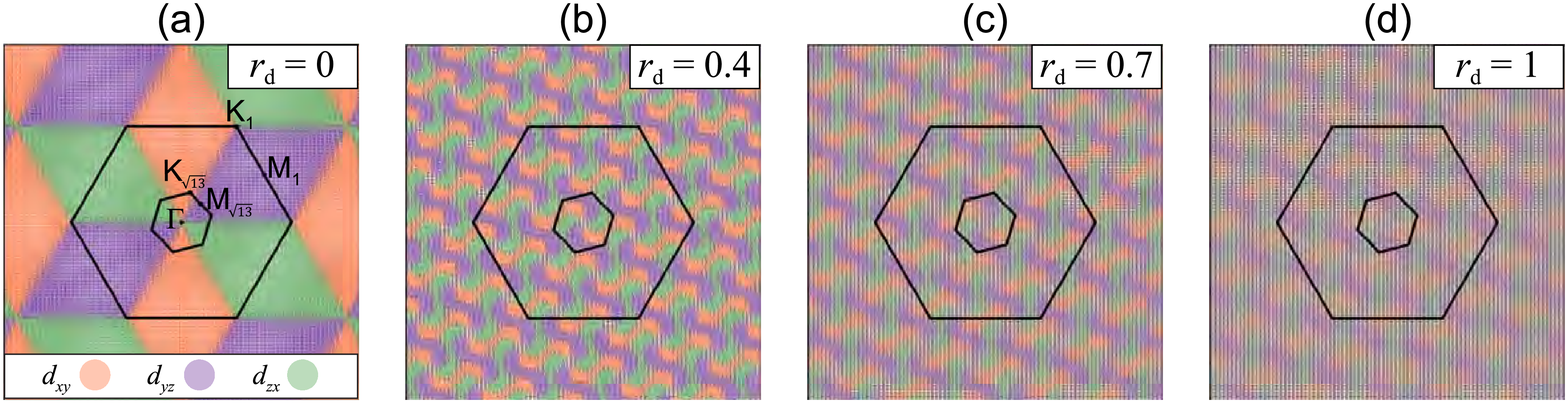}
\caption{Ta $d_{xy}$, $d_{yz}$, and $d_{zx}$ orbital characters of the state crossing $E_{\rm F}$, obtained from (a) $r_d$ = 0 (1${\times}$1 structure), (b) $r_d$ = 0.4, (c) $r_d$ = 0.7, and (d) $r_d$ = 1 (DS structure). The $d_{xy}$, $d_{yz}$, and $d_{zx}$ orbital components are mapped by using red, purple, and green color channels with their brightness, respectively. It is seen that, as $r_d$ increases, the mixing of the three components are enhanced, indicating that the clear ${\bf k}$-dependent Ta $d_{xy}$, $d_{yz}$, and $d_{zx}$ orbital components in $r_d$ = 0 easily hybridize with each other as the lattice distortion increases. The Billouin zones of the 1${\times}$1 and ${\sqrt{13}}{\times}{\sqrt{13}}$ structures are drawn together.}
\label{fig:S1}
\end{figure}

\vspace{1.4cm}
{\bf \large 2. $d$-orbitals projected band structures with respect to the lattice distortion}

\begin{figure}[h!b]
\includegraphics[width=14cm]{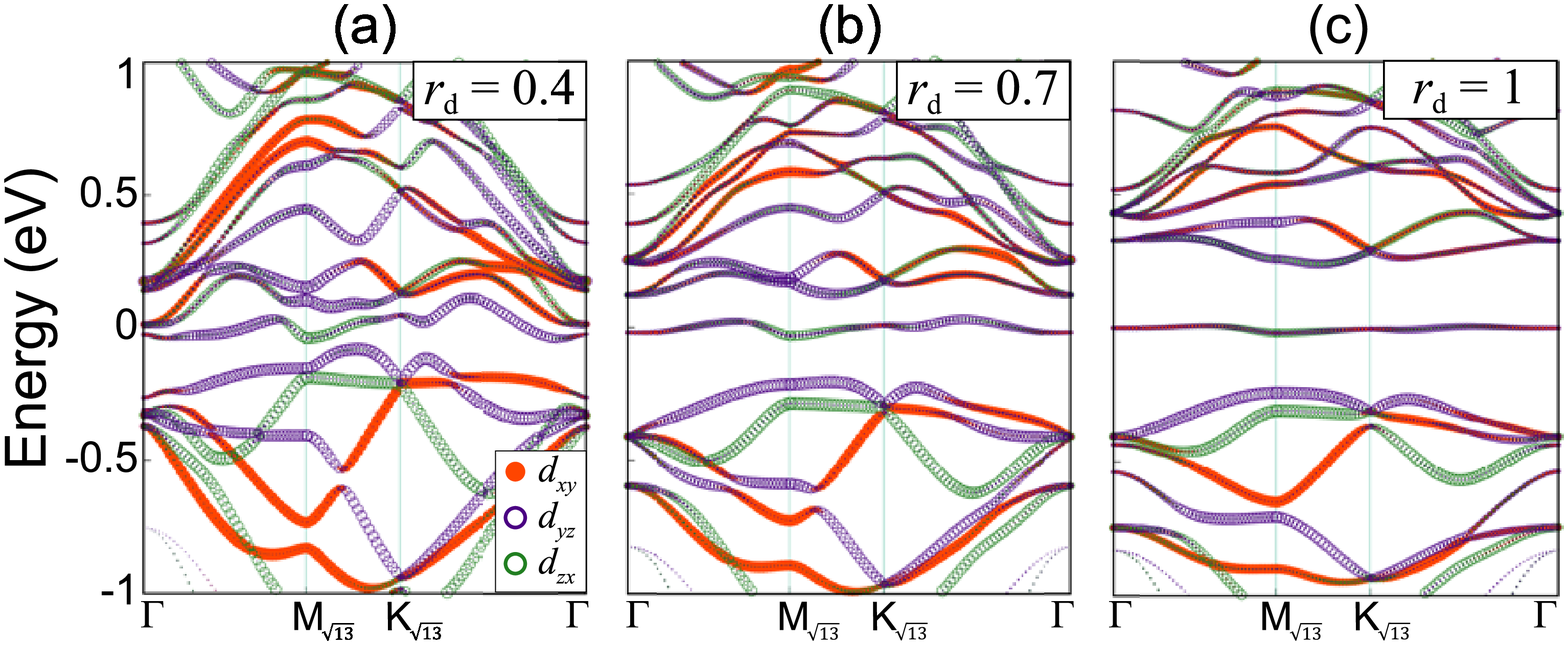}
\caption{Calculated $d$-orbitals projected band structures of 1$T$-TaS$_2$ with respect to $r_{\rm d}$. The bands projected onto Ta $d_{xy}$, $d_{yz}$, and $d_{zx}$ orbitals are drawn with red, purple, and green circles, respectively. Here, the radii of the circle are proportional to the weights of the corresponding orbitals. The energy zero represents the Fermi level $E_{\rm F}$. It is seen that the hybridization of the Ta $d_{xy}$, $d_{yz}$, and $d_{zx}$ orbitals increases with increasing the lattice distortion.}
\label{fig:S2}
\end{figure}

\vspace{1.4cm}
{\bf \large 3. Charge characters of the QMOs with respect to the lattice distortion}

\begin{figure}[h]
\includegraphics[width=14cm]{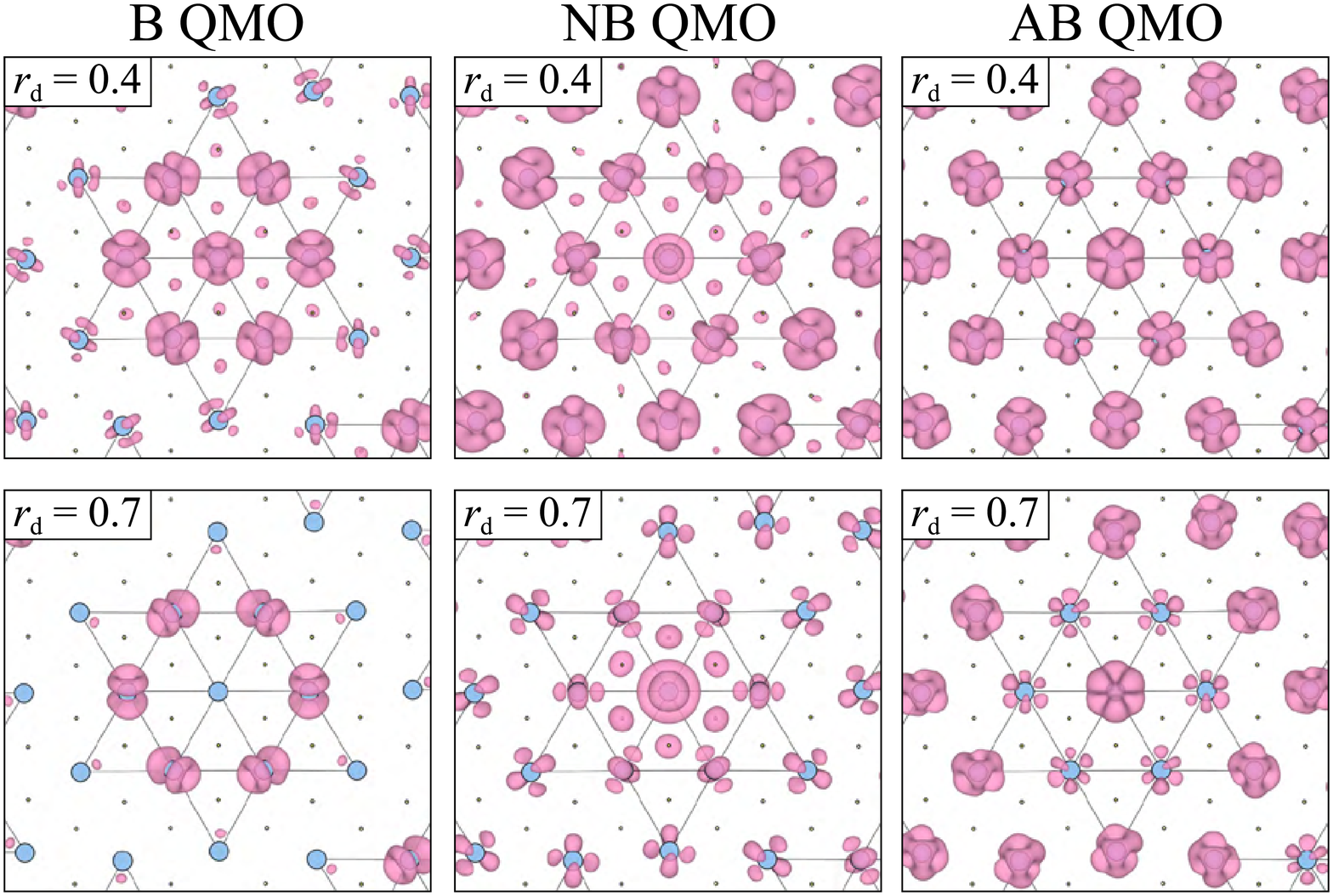}
\caption{Charge characters of the B, NB, and AB QMOs at $r_d$ = 0.4 and 0.7. As $r_d$ increases, the charge characters of the B, NB, and AB QMOs become apparent, indicating that the lattice distortion-induced hybridization of the $t_{\rm 2g}$ states produces the B, NB, and AB QMOs.}
\label{fig:S3}
\end{figure}

\vspace{1.4cm}
{\bf \large 4. Charge and spin characters of an isolated DS molecule}

\begin{figure}[h]
\includegraphics[width=12cm]{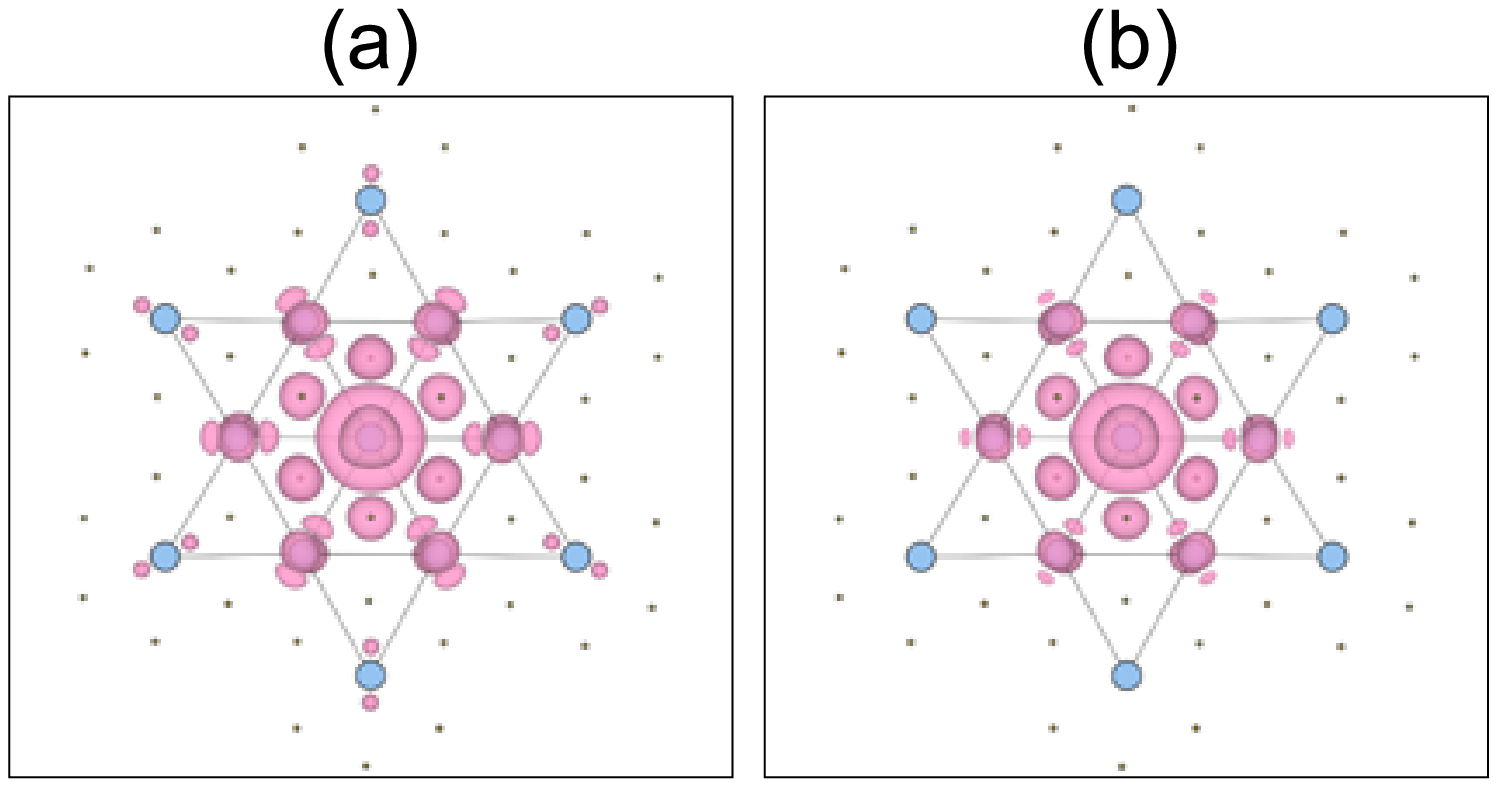}
\caption{(a) Calculated charge distribution of the non-spin-polarized HOMO state of an isolated DS molecule. This charge distribution is almost the same as that of the NB QMO in ML 1$T$-TaS$_2$. The charge density is drawn with an isosurface of 0.003 $e$/{\AA}$^3$. The spin density distribution of the spin-polarized ground state of an isolated DS molecule is given in (b). The spin density is drawn with an isosurface of 0.0015 $e$/{\AA}$^3$. The S atoms at the boundary of an isolated DS molecules are passivated by pseudohydrogen atoms with $\frac{2}{3}$$e$. }
\label{fig:S4}
\end{figure}

\newpage

\vspace{1.4cm}
{\bf \large 5. Calculated band structure of ML 1$T$-TaS$_2$ obtained using the HSE functional}

\begin{figure}[h]
\includegraphics[width=8cm]{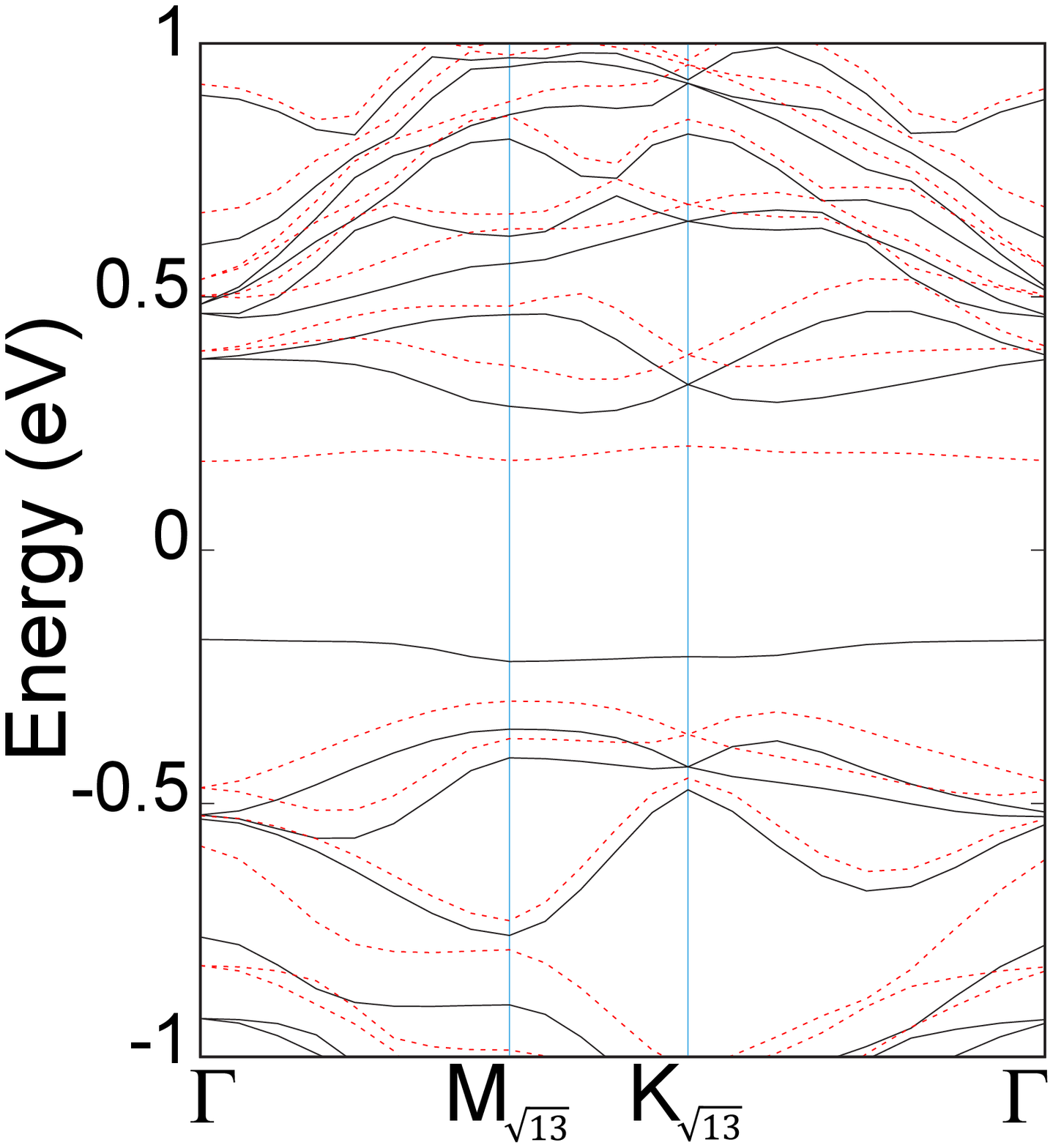}
\caption{Calculated band structure of ML 1$T$-TaS$_2$ obtained using the HSE functional with a mixing factor of ${\alpha}$ = 0.125. The majority and minority bands are drawn with the solid and dashed lines, respectively. }
\label{fig:S5}
\end{figure}

%\newpage

\vspace{1.4cm}
{\bf \large 6. Calculated band structure of the AFM state in ML 1$T$-TaS$_2$}

\begin{figure}[h]
\includegraphics[width=8cm]{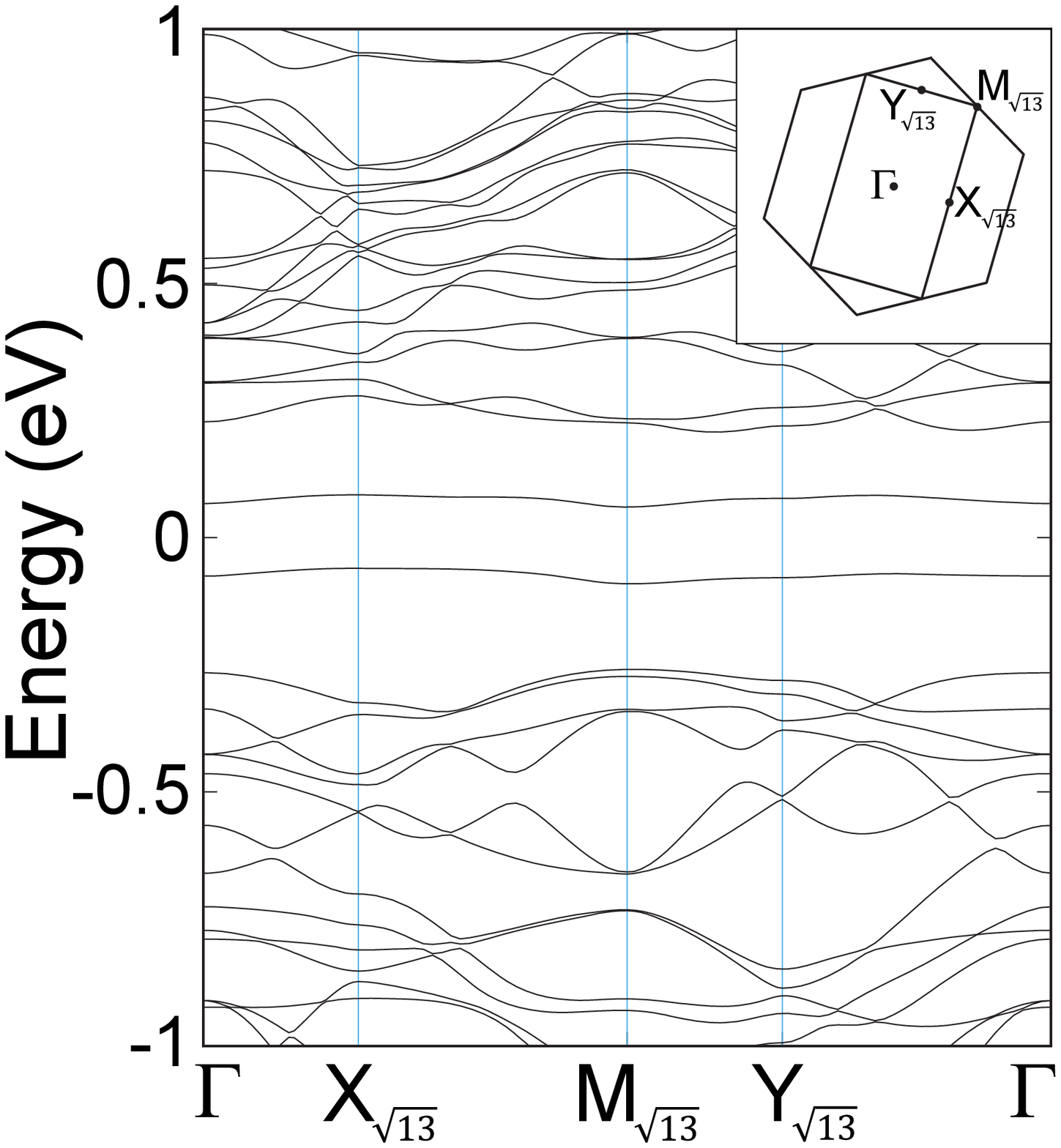}
\caption{Calculated band structure of the AFM state in ML 1$T$-TaS$_2$, obtained using the PBE functional. The band gap is found to be 0.12 eV.}
\label{fig:S4}
\end{figure}

\end{document}